\documentclass[a4paper,twoside,10pt]{article}
\usepackage{graphicx,publaob,bm}

\def\papertype{Contributed paper}
\begin{document}

% Title
\title{On the Numerical Computability of Asteroidal Lyapunov Times}
% Authors
\authors{Enrico Gerlach}

% Addresses and e-mails
\address{$^1$Lohrmann Observatory, Technical University Dresden, Germany}
\Email{enrico.gerlach}{tu-dresden}{de}

% Running titles
\markboth{On the Numerical Computability of Asteroidal Lyapunov Times}{E. Gerlach}

\abstract{To characterize the dynamical stability of an asteroid orbit, the calculation of its Lyapunov-Time $T_L$ is a widely used mean in celestial mechanics. In the present work we investigated the effects of the used computer hardware and integration method on the outcome of such stability computations. We showed that for some asteroids the change of the employed numerical method can change the obtained $T_L$ significantly.

As a result of our investigations we introduced the computability index $\kappa$ as a measure of repeatability of such computations.}

\section{Introduction}

Chaos indicators, like the Lyapunov exponent $\lambda$, are widely used in celestial mechanics to characterize the dynamical behavior of bodies. The principal concept is to calculate the local exponential divergence of arbitrarily close initial conditions. Out of it the stability of their orbits can be determined. One might assume that a numerical calculation of $\lambda$ from the variational equations is straight forward. However, in the literature a lot of discrepancies between different studies dedicated to the same object can be found. 

In this work the dependency of such calculations on hardware and integration methods as a possible source of these differences is investigated. Being part of some bigger project we used here the \emph{MEGNO} indicator to obtain $\lambda$. A paper showing the sensitivity of $T_L$ when being calculated from $\lambda$ directly is in preparation.

\section{Numerical Investigations}

\subsection{The \emph{MEGNO} chaos indicator}

The \emph{MEGNO} chaos indicator $Y$ was defined for a dynamical system $\bm{\dot y} = \bm f(\bm y)$ of dimension $d$ by Cincotta \emph{et al.} (2003) as
\begin{equation}
 Y(t) = \frac{2}{t}\int_0^t \frac{\dot \delta(s)}{\delta(s)}s\: \mathrm{d}s.
\end{equation}
Here $t$ denotes the overall integration time. $\delta$ is the norm of the vector $\bm \delta$ in tangent space. Its time derivative $\bm {\dot \delta}$ satisfies the variational equation
\begin{equation}
 \dot \delta_i =\sum_{k=1}^d \frac{\partial f_i(\bm y)}{\partial y_k}\delta_k.
\end{equation}

The Lyapunov exponent $\lambda$ can be recovered from $Y(t)$ by a least square fit, having the advantage of using the whole dynamical information contained in the integration interval. The Lyapunov time $T_L$ is defined as the inverse of $\lambda$.

\subsection{Set-Up of the Investigation}

For this work a sample of real asteroids was used. These asteroids were chosen in a way, so that their Lyapunov times could be compared to already published values. Additionally, their Lyapunov exponent should be large enough to converge within the integration time span of $10^6$ years to a finite value. So we decided for a sample of 3 main belt asteroids:
\begin{itemize}
 \item Asteroid Helga ($a=3.633$~AU and $e=0.074$), whose Lyapunov time is given around 7000 years in all publications and
 \item Asteroids 1981EY39 ($a=2.238$~AU and $e=0.095$) and 1994EL ($a=2.324$~AU and $e=0.160$), whose Lyapunov times were considered in Kne\v zevic, Z. \& Ninkovic, S. (2005) as difficult to determine.
\end{itemize}

These asteroids were integrated together with a different number of planets of our solar system. All initial orbital elements were taken from the JPL HORIZONS system for the 01/01/2000.

% Table
\begin{table} [hb]
\begin{center}
\caption{\label{tab:lyaptimes}Lyapunov times in $10^3$ years for different integrations. 5 planets refer to the planets Mars to Neptune, while for 7 planets additionally Venus and Earth were integrated together with the asteroids.}
\begin{tabular} {l l r@{.}l r@{.}l r@{.}l r@{.}l}
\hline
 & & \multicolumn{4}{c} {5 planets} & \multicolumn{4}{c} {7 planets}\\
 & & \multicolumn{2}{c} {Intel} & \multicolumn{2}{c} {SGI} &
\multicolumn{2}{c} {Intel} & \multicolumn{2}{c} {SGI}\\ \hline
           & \emph{ODEX}   & 6&69  & 5&77  & 5&87  & 5&98  \\
 Helga     & \emph{SABA} 6\hspace{2ex} & 10&51 & 6&78\hspace{2ex}  & 6&04  & 7&98  \\
           & \emph{SABA} 8 & 9&09  & 7&50  & 6&51  & 7&45\\[1ex]
           & \emph{ODEX}   & 27&73 & 29&59 & 54&63 & 116&80\\
 1981EY39  & \emph{SABA} 6 & 42&97 & 52&62 & 38&28 & 42&56 \\
           & \emph{SABA} 8 & 40&57 & 33&25 & 73&07 & 45&27 \\[1ex]
           & \emph{ODEX}   & 57&95 & 198&20& 49&31 & 36&55 \\
 1994EL    & \emph{SABA} 6 & 55&98 & 74&14 & 24&88 & 53&37 \\
           & \emph{SABA} 8 & 43&59 & 25&87 & 34&66 & 23&95 \\ \hline
\end{tabular}
\end{center}
\end{table}

The numerical integration of the equations of motion and their variational equations was implemented in FORTRAN. As integrators the general purpose extra\-polation scheme \emph{ODEX} from Hairer \emph{et al.} (1987) and the so called \emph{SABA} symplectic scheme as described in Laskar \& Robutel (2001) of order 6 and 8 were used.

For the calculation we had two different hardware platforms at our disposal: a SGI Origin3800 IRIX cluster with MIPS R120000 processor and a personal computer with an Intel Pentium 4 processor running under Linux.

\section{Results}

The results of all the different integrations are shown in Table \ref{tab:lyaptimes}. One sees, that with the same program and the same initial conditions just e.~g. by changing the used computer hardware, the result can change by a factor of 2 and more.

The reasons for these differences are inevitable numerical errors, like round-off and approximation errors, in the course of integration. These errors give rise to a slightly different trajectory, when changing the hardware architecture. And due to the long integration times to reach convergence of $\lambda$, the same initial conditions could end up in very different regions of phase space at the end of the integration.

\begin{figure} [hb]
\includegraphics[width=0.49\textwidth]{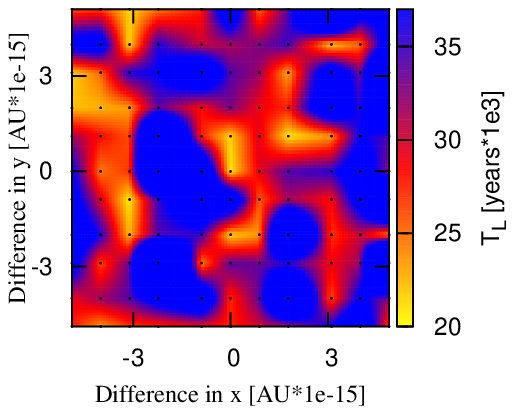} \hfill
\includegraphics[width=0.49\textwidth]{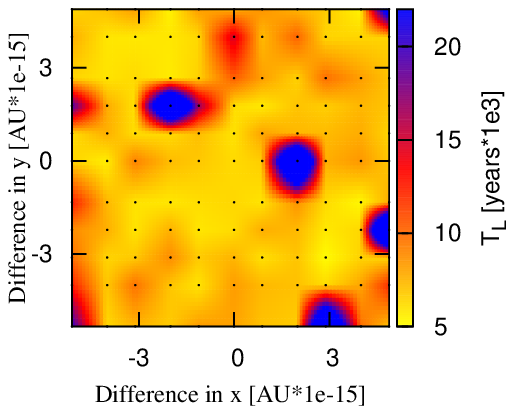}
\caption{\label{fig:eps}Color-coded stability map of the Lyapunov time $T_L$ in the vicinity of asteroid 1981EY39 (left) and Helga (right). The asteroids were integrated together with the planets Mars to Neptune for $10^6$ years using \emph{ODEX}. The initial cartesian $x$ and $y$ coordinates of these asteroids (point of origin) were changed in the last digit to produce a grid of $11\times 11$ initial conditions of test particles, which are shown by black dots. To achieve a smooth representation the area between these points was linear interpolated.}
\end{figure}

By exploring the immediate neighborhood of the initial conditions, we found that the state of the nearby phase space determines the computability of $T_L$. If the phase space is homogeneous, slight changes in the initial conditions e.~g. due to different rounding on different computers will still result in a similar Lyapunov time. For a coarse phase space (caused by overlapping resonances) these differences in rounding could result in very different trajectories of the asteroid and therefore different $T_L$ as well. As an example the phase spaces in the plane of initial $x$ and $y$ coordinate for asteroid Helga and 1981EY39 are given in Figure \ref{fig:eps}.

Furthermore it was found that the general structure of the phase space is robust to a change of the used hardware platform as well as of the integration method. 

\section{The Computability Index $\kappa$}

The information about the nearby phase space can be used to determine the reproducibility of a calculated Lyapunov time  when changing the way of computation. To do so we generated a $5\times 5$ grid of test particles by changing the last digit of the initial $x$ and $y$ coordinate of the asteroid. For these 25 points the corresponding Lyapunov times were calculated. Since the obtained results are not normally distributed, we used a method of robust statistics (the trimmed mean) to calculate the mean Lyapunov time $\overline T_L$ and its standard deviation $\sigma_{\overline T_L}$.

As a measure of reproducibility we defined a computability index $\kappa$ as
\begin{equation}
\kappa = 1-\frac{\sigma_{\overline T_L}}{\overline T_L}.
\end{equation} 

The closer this value is to 1, the more reliable is the result. The computability indexes for the investigated asteroids are shown in Table \ref{tab:comp_index}.

\begin{table}
\begin{center}
\caption{\label{tab:comp_index}The computability index $\kappa$ for the asteroids Helga, 1981EY39 and 1994EL.}
\begin{tabular} {l c c c c}
  \hline
  &\multicolumn{2}{c} {5 planets} & \multicolumn{2}{c} {7 planets} \\ 
  & \hspace{1ex}Intel\hspace{1ex} & \hspace{1ex}SGI\hspace{1ex} & \hspace{1ex}Intel\hspace{1ex} & \hspace{1ex}SGI\hspace{1ex} \\ \hline
  Helga    & 0.95 & 0.97 & 0.95 & 0.97\\ 
  1981EY39 & 0.95 & 0.96 & 0.72 & 0.79\\
  1994EL   & 0.87 & 0.71 & 0.93 & 0.94\\ \hline
\end{tabular}
\end{center}
\end{table}

\section{Conclusions}

We showed that the used computer as well as the chosen integration method can influence the obtained Lyapunov time significantly. With the proposed computability index $\kappa$ the reliability of the stability time of a specific asteroid can be easily estimated.

Finally we would like to remark that the commonly used tests to ensure the quality of an integration -the check of conservation of first integrals- give no indication of the just described problems. The error in energy stayed below $10^{-11}$ for all runs.

% References
\references
Cincotta, P.~M., Giordano, C.~M., Sim{\'o}, C. : 2003, \journal{Physica D}, \vol{182}, 151.

Hairer, E., N\o rsett, S.~P. and Wanner, G. : 1987, \journal{Solving Ordinary Differential Equations I. Non\-stiff Problems}, Springer -- Verlag.

Kne\v zevic, Z., Ninkovic, S. : 2005, \journal{Proceedings of IAU Colloq.197: Dynamics of Populations of Planetary Systems}, 187.

Laskar, J., Robutel, P. : 2001, \journal{Celest.Mech.Dyn.Astr.}, \vol{80}, 39.
\endreferences

\end{document}